# ARTICLE INFORMATION

**Article title**

Gotham Dataset 2025: A Reproducible Large-Scale IoT Network Dataset for Intrusion Detection and Security Research

**Authors**

Othmane Belarbi [a,*], Theodoros Spyridopoulos [a], Eirini Anthi [a], Omer Rana [a], Pietro Carnelli [b], Aftab Khan [b]

**Affiliations**

[a] Cardiff University, School of Computer Science & Informatics, Cardiff, UK

[b] Toshiba Europe Limited, Bristol Research & Innovation Laboratory, Bristol, UK

**Corresponding author's email address and Twitter handle**

BelarbiO@cardiff.ac.uk

**Abstract**

In this paper, a dataset of IoT network traffic is presented. Our dataset was generated by utilising the Gotham testbed, an emulated large-scale Internet of Things (IoT) network designed to provide a realistic and heterogeneous environment for network security research. The testbed includes 78 emulated IoT devices operating on various protocols, including MQTT, CoAP, and RTSP. Network traffic was captured in Packet Capture (PCAP) format using *tcpdump*, and both benign and malicious traffic were recorded. Malicious traffic was generated through scripted attacks, covering a variety of attack types, such as Denial of Service (DoS), Telnet Brute Force, Network Scanning, CoAP Amplification, and various stages of Command and Control (C&C) communication. The data were subsequently processed in Python for feature extraction using the *Tshark* tool, and the resulting data was converted to Comma Separated Values (CSV) format and labelled. The data repository includes the raw network traffic in PCAP format and the processed labelled data in CSV format.

Our dataset was collected in a distributed manner, where network traffic was captured separately for each IoT device at the interface between the IoT gateway and the device. Our dataset was collected in a distributed manner, where network traffic was separately captured for each IoT device at the interface between the IoT gateway and the device. With its diverse traffic patterns and attack scenarios, this dataset provides a valuable resource for developing Intrusion Detection Systems and security mechanisms tailored to complex, large-scale IoT environments. The dataset is publicly available at Zenodo.

# SPECIFICATIONS TABLE

| Subject | Computer Science: Computer Networks and Communications |
|---|---|
| Specific subject area | Artificial Intelligence, Cybersecurity, Computer Networks, Large-scale IoT, Smart Cities, Cyber-attacks. |
| Type of data | - Raw network traffic in Packet Capture (PCAP) format.<br>- Processed network packets with features and labels in Comma Separated Values (CSV). |
| Data collection | Network traffic was generated in the Gotham testbed, an emulated large-scale Internet of Things (IoT) network. It contains 78 emulated IoT devices and is equipped with multiple attack scenarios, including DoS, Remote Command Execution, Ingress Tool Transfer, Reporting, Telnet Brute Forcing, Network Scanning, Periodic C&C Communication, Remote Code Execution, and CoAP Amplification Attack.<br><br>Both benign and malicious network packets were captured in PCAP format using *tcpdump*. The network traffic was captured separately for each IoT device at the interface between the IoT gateway and the device.<br><br>The data was processed in Python to extract relevant features, and it was converted into CSV format. The *Tshark* tool was used to extract traffic features from the captured data. |
| Data source location | Institution: Cardiff University, School of Computer Science & Informatics<br>City: Cardiff<br>Country: United Kingdom<br>GPS Coordinates: 51.489177727305965, -3.1786376893951136 |
| Data accessibility | Repository name: GothamDataset2025<br><br>Data identification number: 10.5281/zenodo.14502760<br><br>Direct URL to data: https://zenodo.org/records/14502760 |
| Related research article | None |

# VALUE OF THE DATA

- Large-scale Internet of Things (IoT) networks, such as smart cities, present significant challenges for network security research due to their heterogeneous nature with diverse devices, protocols, and attack surfaces. Most existing datasets in the field are limited in scale and lack the diversity needed to represent large-scale IoT networks accurately. In addition, most research papers on AI-based Intrusion Detection for IoT networks neglect such networks' heterogeneity, resulting in unrealistic results.

- The proposed dataset addresses these limitations by emulating a large-scale IoT network using the Gotham testbed, offering a realistic environment for network security research. It includes normal traffic generated with protocols such as MQTT, CoAP, and RTSP, alongside diverse attack scenarios, including port scanning, brute force, and DoS. This enables researchers to study the complexities of security mechanisms tailored to large-scale IoT networks.

- Furthermore, the dataset employs a distributed data collection and organisation approach, where network traffic is captured separately for each IoT device at the interface between the IoT gateway and the device. The traffic is stored in device-specific files to reflect real-world IoT deployments where data is generated and maintained at the edge rather than centralised. This design makes the dataset suitable for research on distributed learning techniques in IoT contexts, such as Federated Learning (FL), by enabling the analysis of distributed IoT traffic without relying on centralised data storage. Additionally, this approach eliminates the need for distributing data to simulate realistic heterogeneous conditions, a mechanism most papers in the field rely on.

- The dataset is fully reproducible and extensible as it relies on the open-source Gotham testbed. Researchers can replicate the experiments, integrate new IoT devices, and expand attack scenarios over time under consistent conditions. Furthermore, the processing and labelling pipeline is shared on GitHub to provide clear and reusable tools for dataset creation.

- Researchers can utilise this dataset to develop intrusion detection systems and defence mechanisms specifically designed for large-scale IoT networks. The dataset includes diverse device behaviours, traffic patterns, and attack scenarios. This heterogeneity reflects real-world challenges in securing IoT systems with diverse configurations and evolving threats.

# BACKGROUND

The motivation behind compiling this dataset originated from the limitations of existing IoT network security datasets, which often fail to capture the evolving and heterogeneous nature of large-scale IoT environments [1, 2]. Most datasets are collected in small to medium-scale environments, lacking the structured topologies and configurations in real-world IoT deployments [1]. Typically, devices in these datasets are connected to a single Local Area Network (LAN), which does not accurately reflect the diverse and distributed nature of large-scale IoT networks [1].

Additionally, many datasets overlook the heterogeneity of large-scale IoT systems, failing to include diverse IoT devices, communication protocols, and traffic patterns necessary to simulate real-world conditions [3]. Furthermore, most IoT datasets are collected at centralised locations, limiting their relevance for studying distributed learning applications. Many studies focusing on distributed AI for intrusion detection rely on datasets that inadequately address heterogeneity, potentially affecting the accuracy and applicability of their findings to large-scale IoT scenarios. Often, researchers have to split these centralised datasets to simulate the heterogeneity of real systems.

The key benefits of this dataset lie in its ability to address the significant gaps present in existing IoT network security datasets. By providing a comprehensive representation of large-scale IoT environments, this dataset enables researchers to study distributed learning applications in a more realistic way. Additionally, its focus on heterogeneity allows for the exploration of complex security mechanisms designed specifically for large-scale IoT networks. By addressing these gaps, this dataset provides a reproducible and extensible resource for researchers focused on developing robust security solutions in large-scale IoT settings.

## DATA DESCRIPTION

The data was collected from the Gotham testbed in PCAP format and subsequently processed into CSV format for ease of processing and analysis. The dataset includes several cybersecurity attacks: DoS, Remote Command Execution, Ingress Tool Transfer, Reporting, Telnet Brute Forcing, Network Scanning, Periodic C&C Communication, Remote Code Execution, and CoAP Amplification Attack. It is a multi-class dataset, where each row represents a network packet and contains 23 features along with a label field. The label field can take 10 different values, including Benign, DoS Attack, Remote Command Execution, Ingress Tool Transfer, Reporting, Telnet Brute Forcing, Network Scanning, Periodic C&C Communication, Remote Code Execution, and CoAP Amplification Attack. The dataset is publicly available on the Zenodo repository [4].

The dataset is organised into a hierarchical structure to support scalability and facilitate distributed learning and decentralised analysis. Fig. 1 shows the hierarchical structure of the data repository.

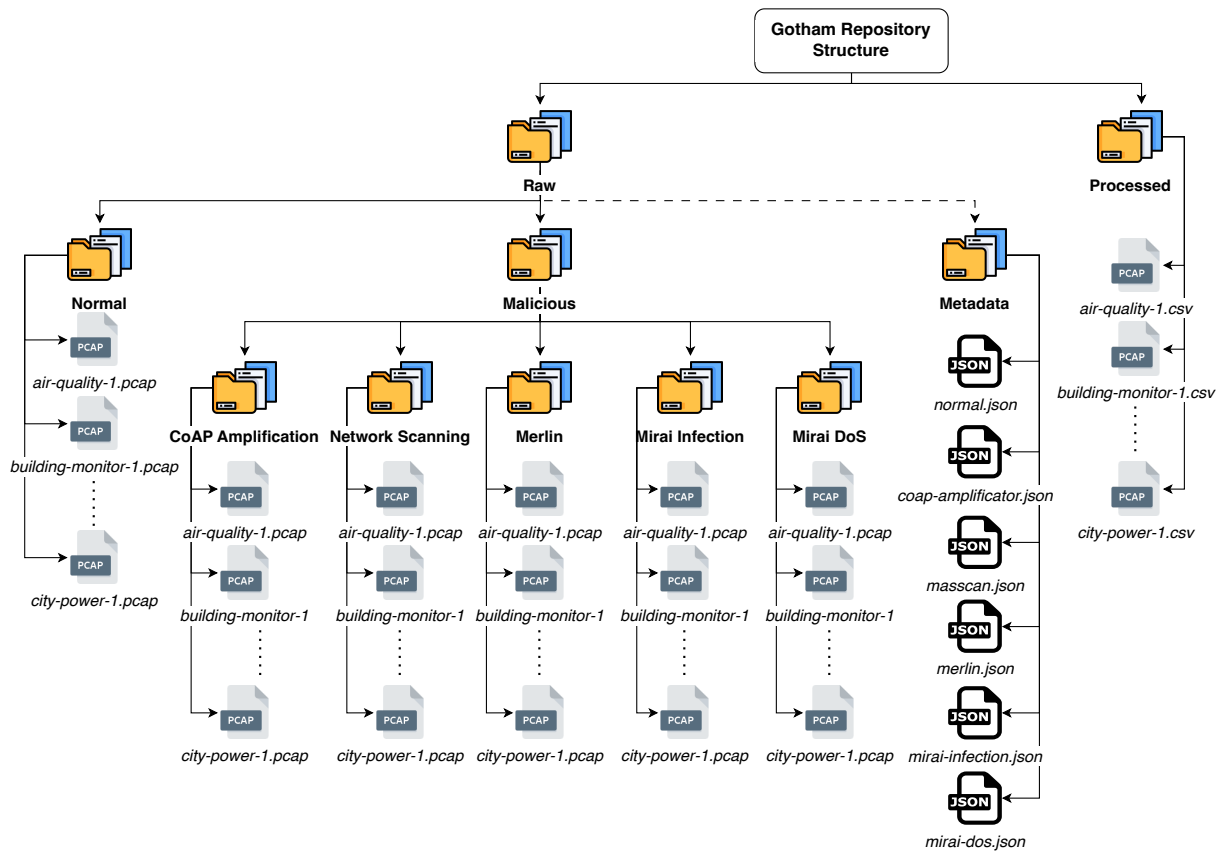

**Fig. 1.** The repository dataset file structure.

Under the root directory, there are two primary folders:

- **Raw Data:** This folder contains the original PCAP files of network traces collected using *tcpdump*. Each PCAP file corresponds to the network traffic for a specific IoT device during a given scenario. Subfolders are categorised by scenario:
  - *Benign:* Contains PCAP files representing normal traffic.
  - *Malicious:* Contains PCAP files categorised by attack type (e.g., Mirai DoS, CoAP amplification).

  A dedicated metadata folder includes files used for labelling the network traffic. The metadata provides contextual information, such as device IP addresses, timestamps, and scenario descriptions, ensuring accurate and reproducible labelling.

- **Processed Data:** This folder contains CSV files derived from the Raw Data. Each CSV file includes feature vectors extracted from network packets, converting unstructured packet data into a structured format ready for machine learning or statistical analysis.

During each attack event, metadata information such as start and end times, source and destination IP addresses, and port details are stored in a log file. For example, a port-scanning attack includes a specified start and stop time, allowing researchers to label individual packets within this time window. Labels for individual packets can thus be inferred from the metadata. The inclusion of a metadata folder provides researchers with the flexibility to develop their own labelling mechanisms based on the metadata. Table 1 provides an overview of the dataset structure, including file names, paths, descriptions, and sizes.

**Table 1.** The dataset public repository file summary.

| File Path | Filename | Description | Size |
|---|---|---|---|
| ./raw/normal/ | iotsim-*<iot-device>*.pcap | These files contain benign network traffic captured over a 2h period with no security attacks. It provides a reference for benign communication. | 7.1GB |
| ./raw/malicious/coap-amplification/ | iotsim-*<iot-device>*.pcap | These files capture network traffic over a 60-minute period, with a CoAP amplification. | 70.6MB |
| ./raw/malicious/network-scanning/ | iotsim-*<iot-device>*.pcap | These files capture network traffic over a 20-minute period, with network scanning attacks. | 496.5MB |
| ./raw/malicious/merlin/ | iotsim-*<iot-device>*.pcap | These files capture network traffic over a 1h30 period,, with Merlin DoS attacks. | 2.6GB |
| ./raw/malicious/mirai-infection/ | iotsim-*<iot-device>*.pcap | These files capture network traffic over a 2h30 period, with Mirai infection attacks. | 3.6GB |
| ./raw/malicious/mirai-dos/ | iotsim-*<iot-device>*.pcap | These files capture network traffic over a 60-minute period, with Mirai DoS attacks. | 11.8GB |
| ./processed/ | iotsim-*<iot-device>*.csv | These files contain a feature-based multi-class labelled dataset, derived from raw traffic data. | 8.86GB |

The network protocol analyser, *Tshark,* is employed to extract features from the recorded network traffic. Several network protocol attributes were analysed for further processing. These protocols include IPv4, ICMP, TCP, UDP, MQTT, CoAP, and RTSP. Each sub-dataset, corresponding to the network traffic for a specific IoT device, contains 23 features, with the final column serving as the label (refer to Table 2).

**Table 2.** The Gotham dataset feature set list.

| No | Feature Name | Prot. Layer | Data Type | Description |
|---|---|---|---|---|
| 1 | frame.time | Frame | Numerical | Arrival Time |
| 2 | frame.len | Frame | Numerical | Frame length on the wire |
| 3 | frame.protocols | Frame | Categorical | Protocols in frame |
| 4 | eth.src | Ethernet | Categorical | Source |

| 5 | eth.dst | Ethernet | Categorical | Destination |
|---|---|---|---|---|
| 6 | ip.dst | IP | Categorical | Destination Address |
| 7 | ip.src | IP | Categorical | Source Address |
| 8 | ip.flags | IP | Categorical | Flags |
| 9 | ip.ttl | IP | Numerical | Time to Live |
| 10 | ip.proto | IP | Categorical | Protocol |
| 11 | ip.checksum | IP | Numerical | Header Checksum |
| 12 | ip.tos | IP | Numerical | Type of Service |
| 13 | tcp.srcport | TCP | Numerical | Source Port |
| 14 | tcp.dstport | TCP | Numerical | Destination Port |
| 15 | tcp.flags | TCP | Categorical | Flags |
| 16 | tcp.window_size_value | TCP | Numerical | Window |
| 17 | tcp.window_size_scalefactor | TCP | Numerical | Window size scaling factor |
| 18 | tcp.checksum | TCP | Numerical | Checksum |
| 19 | tcp.options | TCP | Categorical | TCP Options |
| 20 | tcp.pdu.size | TCP | Numerical | PDU Size |
| 21 | udp.srcport | UDP | Numerical | Source Port |
| 22 | udp.dstport | UDP | Numerical | Destination Port |
| 23 | label | / | Categorical | Label for classification |

The Gotham network topology encompasses various IoT devices. These devices are distributed across four distinct network segments, with each segment representing a different IoT application. Further details on the network topology will be provided in the next section. Table 3 summarises the device types, the application protocols, and associated network traffic statistics.

**Table 3.** The Gotham dataset feature set list.

| Device Type | Application Protocol | # Instance | # Packets | Total Bytes |
|---|---|---|---|---|
| Air Quality | MQTT | 1 | 2,746 | 210,084 |
| Building Monitor | MQTT | 5 | 12,351 | 1,057,100 |
| City Power | CoAP:v1 | 1 | 1,360 | 74,943 |
| Combined Cycle | CoAP:v1 | 10 | 21,808 | 1,223,808 |
| Combined Cycle TLS | CoAP:v1 | 5 | 19,113 | 1,379,967 |
| Cooler Motor | MQTT | 15 | 334,726 | 38,118,306 |
| Domotic Monitor | MQTT | 5 | 13,862 | 1,033,255 |
| Hydraulic System | MQTT | 15 | 72,282 | 18,513,990 |
| IP Camera Street | RTSP | 2 | 752,553 | 1,053,241,255 |
| IP Camera Museum | RTSP | 2 | 1,196,982 | 1,675,390,468 |
| Stream Consumer | RTSP | 2 | 448,460 | 623,285,135 |
| Predictive Maintenance | MQTT | 15 | 257,761 | 38,937,348 |
| **Total** | | **78** | **3,134,004** | **3,452,465,659** |

The network topology consists of 78 IoT devices using multiple application protocols, including MQTT, CoAP, and RTSP. MQTT devices generate moderate network traffic, while CoAP devices produce smaller volumes of traffic. In contrast, RTSP devices like *IP Camera Museum* exhibit significantly higher network activity, with packet counts exceeding 1 million. In total, the dataset contains 3,134,004 packets and 3,452,465,659 bytes of network traffic.

IoT devices transmit data based on specific behaviours. We distinguish between two modes of data transmission and two modes of transmission periodicity: *Open-close* and *Always-open* for transmission modes, and *Continuous* and *Intermittent* for periodicity. In the *Open-close* mode, the device establishes a new connection with the cloud each time it needs to send telemetry data, transmits the data, and then closes the connection. In contrast, the *Always-open* mode involves the device opening a single, persistent connection with the cloud at the start and maintaining it by periodically sending telemetry data and keep-alive messages. For transmission periodicity, the *Continuous* mode means the device continuously transmits telemetry data at regular intervals without interruption. In the *Intermittent* mode, the device alternates between active and inactive periods. During active periods, data is transmitted similarly to the Continuous mode, whereas during inactive periods, only background traffic is transmitted, with no telemetry data. Fig. 2 and Fig. 3 illustrate the heterogeneity of the IoT network.

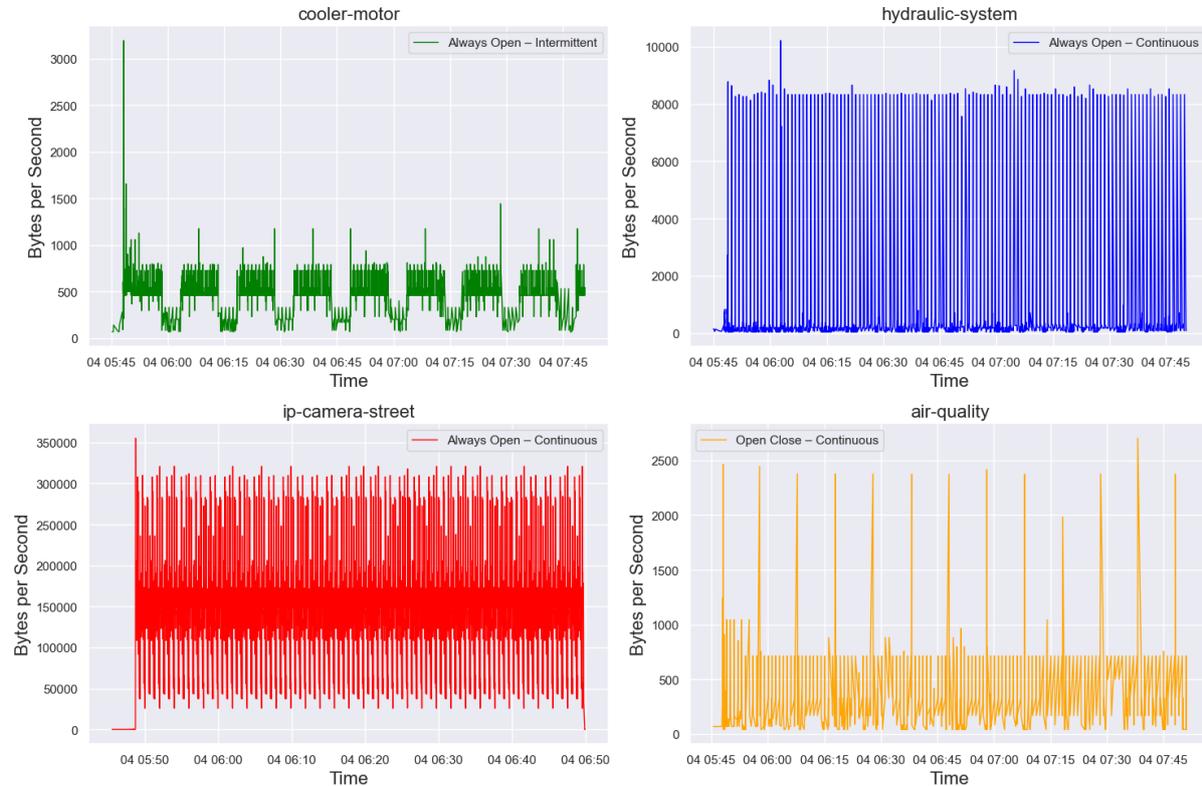

**Fig. 2.** Heterogeneous network behaviours: Always-open – Continuous, Always-open – Intermittent, Open-close – Continuous, Open-close – Intermittent modes.

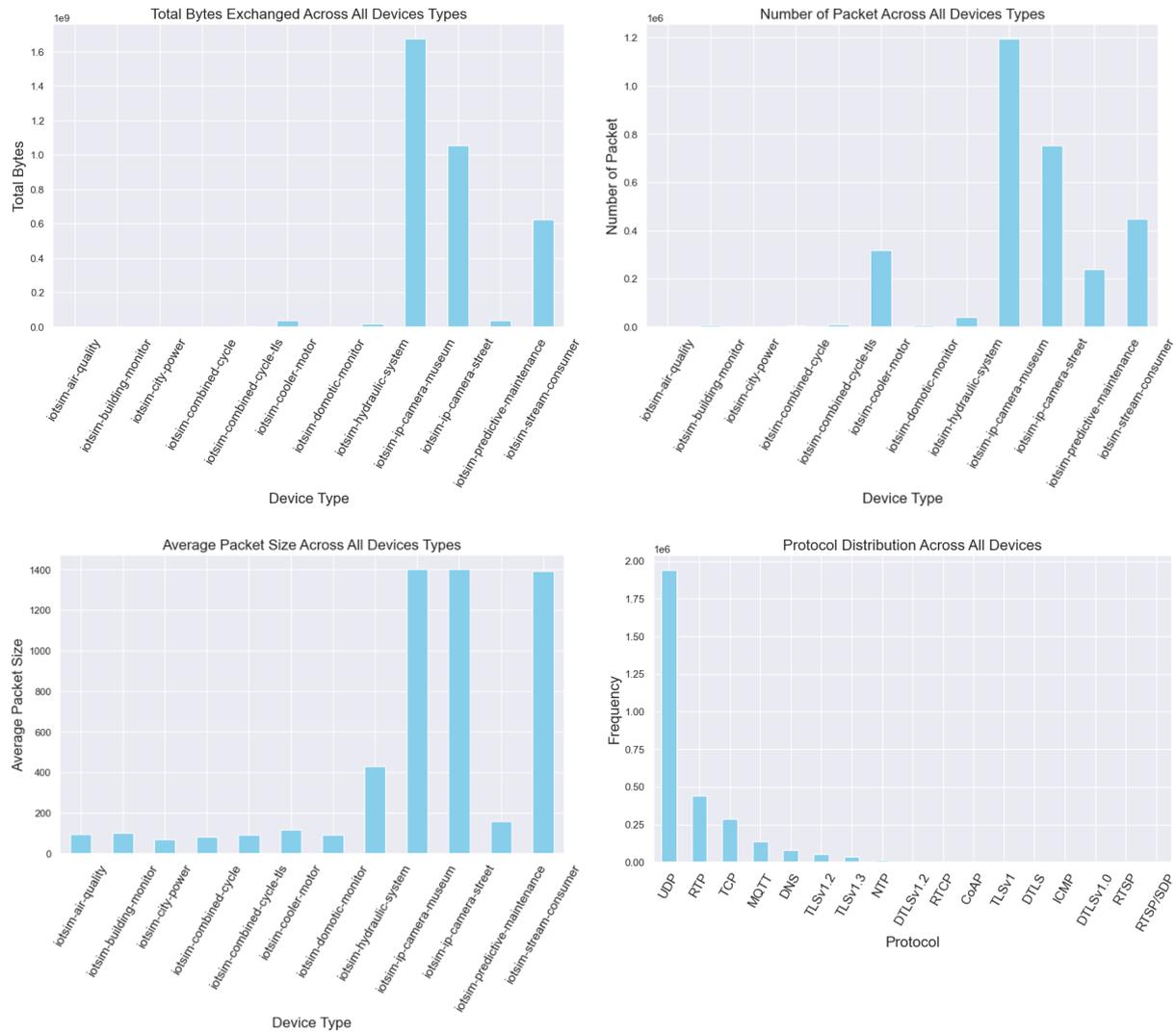

**Fig. 3.** Heterogeneity across IoT devices in terms of packet size, total bytes exchanges, packet volume, and protocol.

The dataset contains various label categories representing both benign and malicious network activities. The Normal label indicates safe traffic with no signs of malicious behaviour. DoS attacks refer to instances where attackers attempt to overwhelm a device or network, disrupting normal functionality. Network Scanning reflects malicious attempts to probe devices for vulnerabilities. Brute Force involves unauthorised attempts to access devices via Telnet by guessing credentials. CoAP Amplification represents DDoS attacks exploiting the CoAP protocol. Additional labels highlight specific stages of attack chains, such as Reporting, where compromised devices send data to a C&C server, Ingress Tool Transfer, which involves the transfer of malicious tools into a network, and File Download, signifying the retrieval of malicious files. The label Periodic C&C Communication indicates ongoing communication between compromised devices and a C&C server.

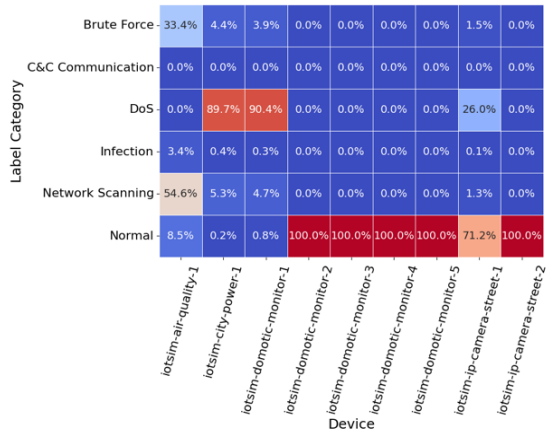 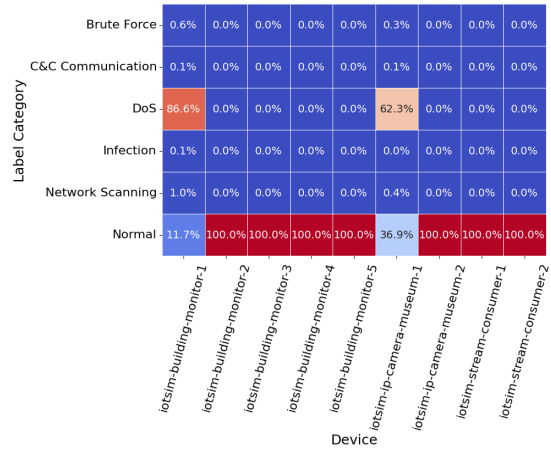

(a) Network segment number 1         (b) network segment number 2

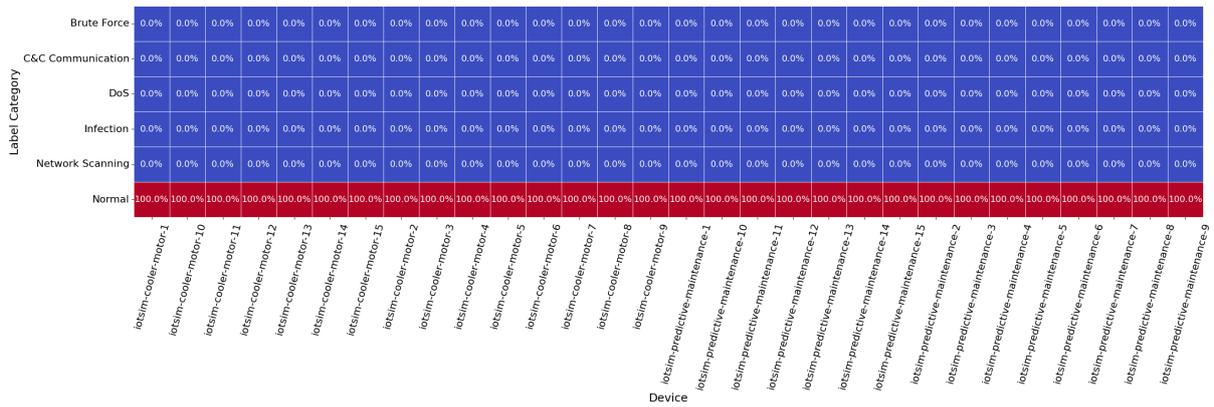

(c) network segment number 3

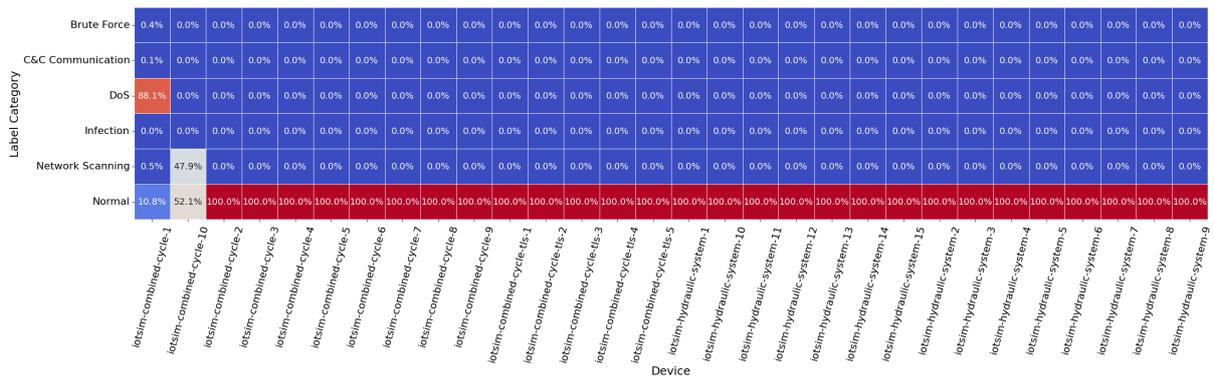

(d) network segment number 4

**Fig. 4.** Label Distribution Across Network Segments Highlighting Dataset Heterogeneity.

Fig. 4 illustrates the label distribution for each device across the four distinct network segments mentioned earlier. Segment 3 contains only benign traffic and represents a secure environment without any attacks. In contrast, the remaining three segments demonstrate varying levels of device compromise, with some devices exhibiting malicious activity while others remain benign. This variability in label distribution reflects the diverse and realistic nature of the dataset, capturing both secure and compromised scenarios. Such heterogeneity is essential to evaluate the performance of intrusion detection systems in large-scale IoT networks within smart cities.

## EXPERIMENTAL DESIGN, MATERIALS AND METHODS

All the experiments were conducted using an AMD Ryzen™ 7 5800H (8-Core, 16-Thread) with 32GB RAM running Ubuntu 24.04 LTS.

**Testbed Overview and Network Structure**

The dataset is suitable for network intrusion detection studies in large-scale IoT environments. The network traffic was acquired in the Gotham testbed Internet of Things [5]. This testbed simulates large-scale IoT environments, leveraging the GNS3 network emulator [6] to provide a realistic setting for IoT network traffic and security research. Gotham comprises a wide array of Docker images and QEMU-based virtual machines emulating IoT/IIoT devices, malware samples, servers, and networking infrastructure, including routers and switches. Real production libraries, Open vSwitch for network switching, VyOS for routing, and various malware samples and red-teaming tools were employed to generate authentic network traffic.

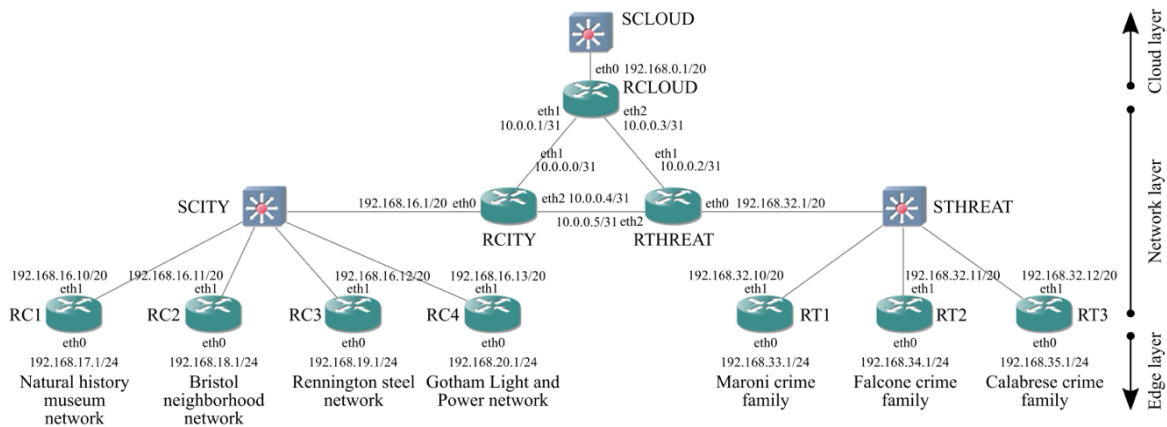

**Fig. 5.** Network diagram of the Gotham testbed [5].

The network topology is segmented into three primary networks – city, cloud, and threat – connected via 10 routers and 30 switches. The city network was further divided into four network segments. Each segment represents a different operational area, with traffic patterns designed to reflect real-world IoT use cases. These segments generate diverse traffic based on typical smart city scenarios, ensuring a realistic simulation of network behaviour. Attack traffic was distributed across these networks to simulate a realistic smart city environment.

**Threat Actors and Attack Scenarios**

The testbed setup involved three distinct threat actors orchestrating a series of attacks: Mirai Malware, Merlin, Port Scanning and Amplification. Each actor employed specific tools and techniques to execute their respective attacks across the network.

*a) Mirai Malware Attacks*

This actor used the Mirai malware [7] to conduct several attack types, with nodes such as the Mirai bot, C&C server, scan listener, loader, and download server adapted from published Mirai source code [8]. Attacks included:

- **Mirai C&C Communication:** Continuous bot-to-C&C communication.
- **Network Scanning:** Mirai bots scanned for open Telnet ports using TCP SYN packets.
- **Brute Forcing:** If a bot found an open Telnet port, it attempted to brute-force credentials with a default IoT credential list.
- **Reporting:** After brute-forcing success, bots reported credentials to the scan listener.
- **Ingress Tool Transfer:** The loader infected vulnerable nodes, downloading and executing Mirai malware.
- **Remote Command Execution:** C&C servers issued commands to bots for various attacks.
- **DoS Attacks:** Included UDP plain attack, DNS attack, TCP ACK, SYN attacks, and others, each executed for 10 seconds.

*b) Merlin-Based Attacks*

This actor leveraged the Merlin cross-platform C&C server [9] with Merlin agents and hping3 for DoS attacks. The C&C communicated through multiple protocols (HTTP/1.1, HTTP/2, HTTP/3, etc.), allowing command execution on victim devices.

- **Merlin C&C Communication:** Periodic contact between Merlin agents and the C&C server.
- **Ingress Tool Transfer:** hping3 was transferred to infected devices for DoS execution.
- **Remote Command Execution:** Commands from the Merlin C&C server triggered attacks.
- **DoS Attacks:** Attacks such as ICMP echo-request, UDP, TCP SYN, and TCP ACK flood generate approximately 5,000 packets at 1 ms intervals.

*c) Port Scanning and Amplification Attacks*

This actor performed extensive network scanning and CoAP amplification attacks using tools like Nmap, Masscan, and AMP-Research.

- **Network-Wide Scanning:** Conducted with Masscan for specific TCP ports at various packet rates (100, 1000, and 10,000 packets/s) and with Nmap for UDP ports.
- **CoAP Amplification Attack:** A CoAP device in the city network was leveraged to amplify attacks for 15 seconds against a target.

The malicious events were conducted in sequential batches, one after the other. The dataset generation workflow, illustrated in Fig. 6, consisted of several stages:

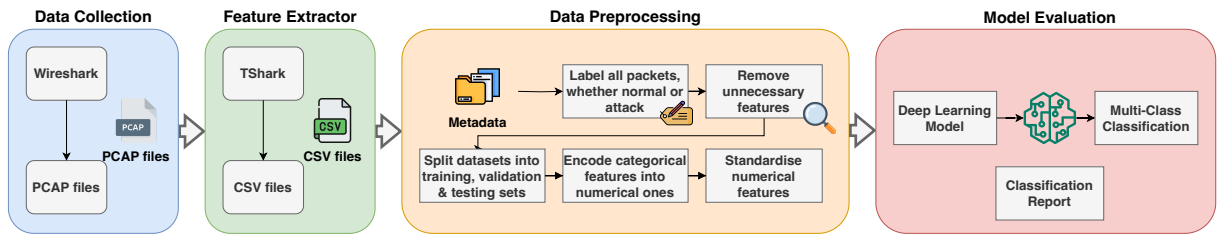

**Fig. 6.** Dataset generation methodology.

**Data Collection and Feature Extraction**

Network packets were captured from IoT devices using *tcpdump* in a distributed manner, generating raw PCAP files. Each PCAP file contains network traffic for a specific IoT device during a given network event, enabling decentralised analysis and distributed learning. The network traffic was captured separately for each IoT device at the interface between the IoT gateway and the device.

First, benign network traffic was captured over 2 hours without any security attacks, providing a baseline for normal communication. This data enables researchers to create baseline models for intrusion detection. Subsequently, various attacks were launched sequentially, with each attack lasting between 1 and 1.5 hours.

Features were then extracted from the PCAP files using *Tshark*. This extraction produced CSV files containing the relevant feature vectors, detailing characteristics such as packet size, timestamp, protocol type, source/destination IP addresses, and ports.

The data generation, collection, and feature extraction processes were automated using Python scripts to ensure reproducibility and flexibility for adjustments.

**Data Labelling**

The extracted data were labelled in Python within a controlled environment, following a structured methodology to accurately distinguish between normal and malicious traffic based on metadata. The metadata was generated during the experimental setup by logging details about the attack scenarios in real-time. It included timestamps of attacks, network packet characteristics, and the attacker's IP address.

A **timing synchronisation strategy** was applied to label network packets in the CSV files. This strategy involved aligning the timestamps of attack phases with the captured network traffic. For each attack phase, packets within the corresponding time window were labelled as malicious, while those outside the attack intervals were labelled as normal. This approach ensures precise labelling, maintains the integrity of the dataset and enables researchers to analyse attack behaviours in relation to the exact timing of network traffic. Fig. 7 provides a simplified overview of the timing synchronisation strategy.

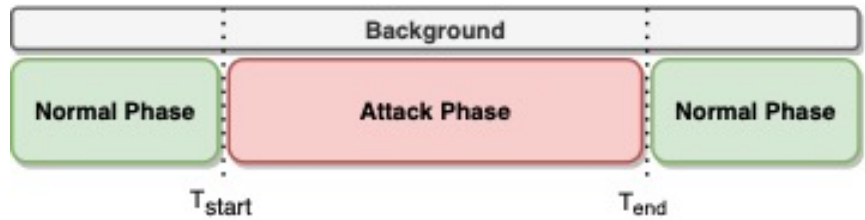

**Fig. 7.** An example of the timing synchronisation labelling strategy.

The labelling process followed these steps:

- **Traffic Filtering:** Raw network packets were filtered to include only IPv4/IPv6 packets where the IoT device IP address was the source or destination, isolating interactions directly involving the IoT devices.
- **Normal Traffic Labelling:** Initially, traffic classified as normal, such as device-to-server communication, was labelled.
- **Malicious Traffic Labelling:** Subsequent malicious events were labelled based on specific attack signatures and metadata, covering various stages of the intrusion process.

The output of the methodology is a set of CSV files containing network packets labelled with extracted features. These files can be used to develop Intrusion Detection Systems (IDSs) using Machine Learning (ML) or Deep Learning (DL) models to detect cyber-attacks on IoT devices and large-scale IoT networks. These models can be trained in both centralised and distributed learning settings.

The data processing scripts are available on GitHub at https://github.com/othmbela/gotham-network-packet-labeller to ensure reproducibility. Researchers can use these scripts to process raw network traffic and create their datasets. The repository includes instructions and configurations for easy use and customisation.

**Use Examples**

To demonstrate how the dataset can be utilised for intrusion detection research, we present an evaluation of a basic Deep Neural Network (DNN) trained on the dataset in an FL setup. A DNN was selected because DL models are particularly effective at capturing complex patterns in high-dimensional network traffic data. Prior studies [10] have evaluated conventional ML models (e.g., Decision Tree (DT), Random Forest (RF), Support Vector Machine (SVM), K-Nearest Neighbors (KNN)) alongside DNNs for cyber-attack detection in FL settings. The experimental results demonstrated the effectiveness of DNNs in handling large-scale and heterogeneous datasets.

FL was selected to align with the decentralised nature of IoT systems, where data is generated across multiple devices rather than in a centralised manner. Beyond decentralisation, FL mitigates privacy risks by ensuring that raw network traffic data remains on each IoT device [11]. This is particularly relevant for real-world smart city deployments, where transmitting large volumes of traffic data to a central server may introduce security risks as well as network congestion [11]. Additionally, this setup also supports scalability, as models can be trained without centralising vast amounts of data, making it ideal for large-scale IoT networks.

In this experiment, we used 10 rounds of FL, which is a typical number used in similar studies [10]. In each round, a subset of 13 IoT devices was sampled for model training. Each IoT device splits its dataset into a training set (80%) and a testing set (20%). This approach ensures that the model is evaluated on data it has not encountered during training. The training process was distributed across the 13 selected devices, allowing each device to locally train the model on its own data and share the model updates with a central aggregator. After aggregating the local updates, the global model was tested on a centralised server using the testing subsets, ensuring consistency in performance assessment across rounds.

**Table 4.** Classification report of DNN.

|  | Precision | Recall | F1-Score |
|---|---|---|---|
| **Brute Force** | 0.67 | 0.54 | 0.60 |
| **C&C Communication** | 0.99 | 0.95 | 0.97 |
| **DoS** | 0.88 | 1.00 | 0.94 |
| **Infection** | 0.00 | 0.00 | 0.00 |
| **Network Scanning** | 0.49 | 0.66 | 0.56 |
| **Normal** | 1.00 | 0.52 | 0.69 |
| **Accuracy** |  |  | 0.89 |
| **Weighted Average** | 0.90 | 0.89 | 0.87 |

The model achieved an accuracy of 89%, as depicted in Table 4. This high accuracy demonstrates the model's ability to effectively differentiate between benign and malicious traffic, even in the context of an FL environment with diverse IoT devices. Nonetheless, the model also exhibited some classification errors, which suggest areas for further improvement, particularly in handling specific attack types or device-specific characteristics.

To facilitate reproducibility, we provide Jupyter Notebooks in the repository that demonstrate how to use the dataset here. These notebooks can be executed in a Google Colab environment and cover data loading, preprocessing, and model training to help researchers extend and refine our approach.

# ETHICS STATEMENT

The authors have read and followed the ethical requirements for publication in Data in Brief and confirm that this work does not involve human subjects, animal experiments, or any data collected from social media platforms.

# CRediT AUTHOR STATEMENT

**Othmane Belarbi:** Conceptualization, Methodology, Software, Data Curation, Writing – original draft. **Theodoros Spyridopoulos:** Methodology, Validation, Writing – review & editing, Supervision. **Eirini Anthi:** Writing – review & editing. **Pietro Carnelli:** Writing – review & editing. **Aftab Khan:** Writing – review & editing.

# ACKNOWLEDGEMENTS

This research did not receive any specific grant from funding agencies in the public, commercial, or not-for-profit sectors.

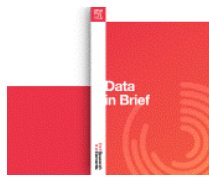

# DECLARATION OF COMPETING INTERESTS

The authors declare that they have no known competing financial interests or personal relationships that could have appeared to influence the work reported in this paper.